\documentclass[aps,twocolumn,prd,preprintnumbers]{revtex4}
\usepackage{graphicx}

\newcommand{\mpl}{m_{\rm Pl}}
\newcommand{\lpl}{l_{\rm Pl}}

\begin{document}

\preprint{FTPI-MINN-05/23}
\preprint{UMN-TH-2409/05}

\title{Inflation: A graceful entrance from Loop Quantum Cosmology}

\author{N. J. Nunes}
\affiliation{School of Physics and Astronomy, University of Minnesota, 116 Church
Street S.E., Minneapolis, Minnesota 55455, USA}
\email{nunes@physics.umn.edu}

\date{\today}

\begin{abstract}
The evolution of a scalar field is explored taking into account the
presence of a background fluid in a positively curved Universe in the
framework of loop quantum cosmology. Though the mechanism that
provides the initial conditions for inflation
extensively studied in the literature, is still available in this
setup, it demands that
 the initial kinetic energy of the field be comparable to the
energy density of the background fluid if the field is initially
situated at the minimum of the potential. It is found, however, that
for potentials with a minimum such as the chaotic inflation model,
there is an additional mechanism 
that can provide the correct initial
conditions for successful inflation even if initially the kinetic
energy of the field is subdominant by many orders of magnitude. In this
latter mechanism the field switches direction when the Universe is
still in the expanding phase. 
The kinetic energy gained while the field rolls down
the potential is subsequently enhanced when the universe enters the
collapsing phase pushing the field one step up the potential. This
behavior is repeated on every
cycle of contraction and expansion of the Universe until the field
becomes dominant and inflation follows.
\end{abstract}


\maketitle

\section{Introduction}
Currently, the leading background independent and non-perturbative
candidate for a quantum theory of gravity is loop quantum gravity \cite{Rovelli:1997yv,Thiemann:2002nj,Corichi:2005bn} which is a canonical quantization of general relativity based in Ashtekar's variables.
This approach provides a discrete structure of geometry. Continuous spacetime emerges in the large eigenvalue limit of quantum geometry. Loop quantum Cosmology (LQC) is the application of loop quantum gravity to homogeneous and isotropic mini-superspaces \cite{Bojowald:2002gz}.
An important feature of LQC is that eigenvalues of the inverse scale
factor operator are proportional to {\it positive} powers of the scale factor below a critical scale $a_*$ \cite{Bojowald:2002ny}.
In particular, it was shown that this property affects the behavior of the kinetic energy of a scalar field at small scales.
There has been considerable interest
in understanding the dynamics of a scalar field in this semi-classical phase, in particular, its role in the origin of inflation
\cite{Bojowald:2002nz,Tsujikawa:2003vr,Bojowald:2004xq,Lidsey:2004ef,Lidsey:2004uz,Date:2004yz,Mulryne:2004va,Vereshchagin:2004uc,Mulryne:2005ef}, avoidance of a big crunch
in a closed Universe \cite{Singh:2003au,Date:2004fj} and non-singular bounces in the cyclic scenario \cite{Bojowald:2004kt}.

In classical standard cosmology the energy density of a perfect fluid with constant equation of state evolves as $\rho \propto a^{-3(w_{\rm cl}+1)}$ hence , diverging as $a \rightarrow 0$ for $w_{\rm cl} > -1$. Consequently, the presence of a matter component into the dynamics may lead to singularities. The behavior of matter in the semi-classical phase,
however, is not completely understood as a theory of quantum gravity including matter is yet to be constructed. A purely phenomenological approach has recently been attempted
\cite{Singh:2005km}.
It was found that similarly to the case of
a scalar field, the classical cosmology equivalent of the energy density of a perfect
fluid is modified in LQC. It varies proportionally to positive powers of the scale factor (for $w_{\rm cl} > 1/4$) at small scales, preventing it from diverging.

The semi-classical phase of LQC is defined for values of the scale factor in the interval
$a_i < a < a_*$, where $a_i^2 \equiv \gamma\lpl^2$ and
$a_*^2 \equiv j \,a_i^2/3$. The Barbero-Immirzi parameter is
$\gamma = \ln 2/\sqrt{3}\pi \approx 0.13$ \footnote{See also \cite{Domagala:2004jt,Meissner:2004ju} where $\gamma = 0.274$. Our results are reproduced with this definition by rescaling $j$ using $\tilde{j} = 0.4649 \,j$.} and
$j$ is a half integer quantization parameter. The evolution equations are modified by the presence of non-perturbative effects.
The discrete nature of spacetime is important below $a_i$ and is described by a difference equation. The evolution equations take the standard
classical form above $a_*$.

The modified Friedmann equation for a positively curved Friedmann-Robertson-Walker Universe sourced by
a scalar field $\phi$ with self-interaction potential $V(\phi)$ and a background fluid
with classical equation of state $w_{\rm cl}$, has the form
\begin{equation}
\label{eqfriedmann}
H^2 \equiv \left(\frac{\dot{a}}{a} \right)^2 = \frac{\kappa^2}{3} \left(\rho_\phi + \rho_{\rm b} \right) - \frac{1}{a^2} \,,
\end{equation}
where we have defined $\kappa^2 \equiv 8 \pi \lpl^2 = 8 \pi \mpl^{-2}$ and the energy densities of the scalar field and background fluid as \cite{Singh:2005km}
\begin{eqnarray}
\rho_\phi &=& \frac{1}{2} \frac{\dot{\phi}^2}{D} + V(\phi) \,, \\
\rho_{\rm b} &=& \rho_0 \, D^{w_{\rm cl}} \, a^{-3(1+w_{\rm cl})} \,.
\end{eqnarray}
The quantum correction function $D(q)$ is defined by \cite{Bojowald:2002nz}
\begin{eqnarray}
D(q) &=& \left(\frac{8}{77}\right)^6 q^{3/2}
\left\{ 7\left[(q+1)^{11/4} - |q-1|^{11/4}\right]
\right. \nonumber \\
&-& \left. 11q\left[(q+1)^{7/4}- {\rm sgn} (q-1)|q-1|^{7/4}
\right]\right\}^6 \,
\end{eqnarray}
with $q \equiv (a/a_*)^2$. In the semi-classical phase ($a \ll a_*$), the quantum correction function varies as $D \propto a^{15}$, has a maximum near $a = a_*$ and falls monotonically to the classical limit $D = 1$ in the classical phase ($a > a_*$).

The equation of motion for the scalar field is given by
\begin{equation}
\label{eqphi}
\ddot{\phi} + 3 H \left( 1- \frac{1}{3} \, \frac{d \ln D}{d \ln a} \right)
\dot{\phi} + D \, \frac{d V}{d \phi}  = 0 \,.
\end{equation}

Differentiating Eq.~(\ref{eqfriedmann}) and substituting for $\ddot{\phi}$
using Eq.~(\ref{eqphi}) gives
\begin{eqnarray}
\label{dotH}
\dot{H} &=& -\frac{\kappa^2}{2} \left(1-\frac{1}{6} \, \frac{d \ln D}{d \ln a} \right)
\nonumber \\ &~&
 -\frac{\kappa^2}{2} \left(1+ w_{\rm cl} -\frac{w_{\rm cl}}{3} \, \frac{d \ln D}{d \ln a} \right) \rho_{\rm b} + \frac{1}{a^2} \,.
\end{eqnarray}

The effective equations of state of the scalar field and background fluid can be written as
\cite{Lidsey:2004ef,Singh:2005km}
\begin{eqnarray}
w_{\phi} &=& -1 + \frac{2\dot{\phi}^2}{\dot{\phi}^2+ 2 D V} \left(1-\frac{1}{6} \, \frac{d \ln D}{d \ln a} \right) \,, \\
w_{\rm b} &=& w_{\rm cl} \left(1-\frac{1}{3} \, \frac{d \ln D}{d \ln a} \right) \,.
\end{eqnarray}
We can see from these equations that deep into the semi classical phase, the equations of
state become $w_{\phi} = -4$ (neglecting the potential) and $w_{\rm b} = -4 w_{\rm cl}$.
It was shown in Ref.~\cite{Singh:2003au} that such a feature allows the energy
density of the scalar field to cancel out the curvature term allowing the Universe to re-bounce at small values of the scale factor, hence avoiding a big crunch. A recollapse also occurs at a large value of the scale factor in the classical phase as the kinetic energy of the field scales more quickly than the curvature term. An oscillatory Universe is therefore the natural outcome of the LQC corrections in a positively curved space sourced by a scalar field. Since the kinetic energy of the field never vanishes while the potential is negligible, it was discussed in Refs.~\cite{Lidsey:2004ef,Bojowald:2004kt,Mulryne:2004va,Mulryne:2005ef}, that the
oscillatory behavior of the Universe enables the scalar field to move up in its potential
establishing the initial conditions for slow roll inflation
(when $\dot{\phi}^2 \approx V$), even admitting that
the field starts its evolution at the minimum of the potential with small kinetic
energy.

In this paper we extend these works by including a background fluid in the dynamics of the Universe. We find that a mechanism similar to the one just described is still a possibility but it only results into an inflationary expansion for a sufficiently large contribution of the field's kinetic energy. Nonetheless, we describe an alternative mechanism through which the field can become dominant, even if the kinetic energy is initially many orders of magnitude below the energy density of the background fluid, and discuss the parameter space where it is most efficient.

Under the notion of perfect fluid, it is typically assumed that
quantum effects play a negligible role in the dynamics of the
corresponding quantum field such that the concept of an equation of state
can be applied. Since in our study we are precisely concerned with
evaluating the effects of loop quantum corrections on a fluid with a
constant classical equation of state, we are in danger of dealing with
an inappropriate definition. We must necessarily assume that the
dynamics affected by the loop quantum modifications has a time scale much
larger than the time necessary to ensure thermodynamical
equilibrium (see also \cite{Banerjee:2005ga,Singh:2005km}). 
This assumption, however, may become invalid when the
scale factor of the Universe becomes close to $a_i$ where the full
quantum theory comes into play. We will see, however, that the interesting and
viable models including a background fluid are only well defined for $a_*
\gg a_i$ and $a \approx a_*$, hence, we expect that the quantum
corrections exert a harmless role where the definition of equation
of state and energy density is concerned.

\section{Critical points and stability}

The equations of motion (\ref{eqphi}) and (\ref{dotH})
can be rewritten as a system of first order differential equations by defining the dimensionless quantities
\begin{eqnarray}
x &=& \frac{\kappa}{\sqrt{6}} \, \frac{\dot{\phi}}{\sqrt{D}} \, a \,, \\
y &=& \frac{\kappa}{\sqrt{3}} \, \sqrt{|V|} \, a \,, \\
z &=& \dot{a} \,, \\
\label{defw}
w &=& \frac{\kappa}{\sqrt{3}} \, \sqrt{|\rho_{\rm b}|} \, a \,.
\end{eqnarray}
For generality, we have considered the possibility that the scalar potential and/or the energy density of the background fluid are negative (e.g. a negative cosmological
constant).
The system is now governed by the following equations:
\begin{eqnarray}
x' &=& 2xz\left(\frac{\Delta}{4}-1\right) \pm \sqrt{\frac{3}{2}D} ~\lambda y^2 \,, \\
y' &=& yz - \sqrt{\frac{3}{2}D} ~\lambda xy \,, \\
z' &=& 2x^2\left(\frac{\Delta}{4} -1 \right) \pm y^2 \nonumber \\
&~&
\pm \frac{1}{2} w^2 \left[w_{\rm cl} (\Delta-3) -1 \right] \,, \\
q' &=& 2qz \,, \\
\lambda' &=& -\sqrt{6 D} \, \lambda^2 x (\Gamma -1) \,,
\end{eqnarray}
subject to the Friedmann constraint
\begin{equation}
\label{friedmann}
x^2 \pm y^2 -z^2 \pm w^2  = 1\,.
\end{equation}
Here a prime corresponds to differentiation with respect to conformal time $d \tau = dt/a$.
We have also used the definitions:
\begin{eqnarray}
\lambda &=& -\frac{1}{\kappa} \, \frac{1}{V} \, \frac{dV}{d \phi} \,, \\
\Gamma &=& V ~\frac{d^2 V}{d \phi^2}~ \left(\frac{d V}{d\phi} \right)^{-2} \,, \\
\Delta &=& 2 \, \frac{d \ln D}{d \ln q} \,.
\end{eqnarray}
It is useful to define the dimensionless quantities
\begin{eqnarray}
\label{defrv}
r = \frac{\kappa^2}{3}\rho_0 \, a_*^{-1-3w_{\rm cl}} \,, \hspace{1cm}
v = \frac{\kappa^2}{3} V a_*^2 \,,
\end{eqnarray}
so that $y$ and $w$ can now be written as
\begin{eqnarray}
\label{y2w2}
\pm y^2 =  v q \,, \hspace{1cm}
\pm w^2 = r D^{w_{\rm cl}} \, q^{-(1+3w_{\rm cl})/2} \,.
\end{eqnarray}
These definitions will allow us to perform a general study of the system
regardless of the value of the quantization parameter $j$ which is now
enclosed in $r$ and $v$.

We will assume for now that the potential is constant, i.e. $\lambda = 0$. In the next section we will extrapolate the results to a varying potential under the assumption that the variation is slow enough such that we can refer to ``instantaneous critical points''.
Following \cite{Mulryne:2005ef} we are looking for static solutions such that $\dot{a} = \ddot{a} = 0$ or, in the current  variables, $q'=0$ and $z' = 0$. The system has, therefore, critical points in $z_c = 0$ and $q_c$ such that
\begin{equation}
\label{critical}
\pm y_c^2 (6-\Delta_c) \pm w_c^2 \, (\Delta_c-3)(w_{\rm cl}-1) + \Delta_c-4 = 0 \,.
\end{equation}
Note that $w_c$ is the value of $w$ defined in Eq.~(\ref{defw}) at the critical point and $w_{\rm cl}$ is the classical value of the equation of state.

We must also guaranty that the kinetic energy of the scalar field is non-negative at these critical points, hence it follows, using the Friedmann constrain (\ref{friedmann}) and Eq.~(\ref{critical}) that
\begin{eqnarray}
\label{kinetic}
x_c^2 = \frac{1}{6-\Delta_c}\left[2 \pm w_c^2 \left( w_{\rm cl}(\Delta_c-3) - 3 \right) \right] \ge 0 \,.
\end{eqnarray}

\begin{figure}[!t]
\includegraphics[width = 8.5cm]{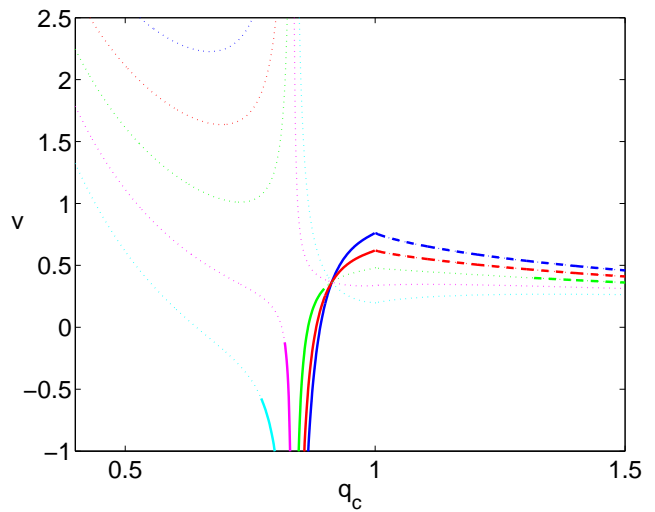}
\caption{\label{sol1} Dependence of $v$ on the critical value $q_c$.
From top to bottom the curves represent $r = 0$, 0.3, 0.6, 0.9 and  1.2. Solid lines represent stable critical points and
dashed lines represent saddle points. Dotted lines correspond to the position of the critical points if condition (\ref{kinetic}) was satisfied.}
\end{figure}

In Fig.~\ref{sol1} we show, by dotted lines,
the relation between $v$ and $q$ of the
system's static solutions as determined form Eq.~(\ref{critical}).
The case studied in Ref.~\cite{Mulryne:2005ef} corresponds to the top lines, $r = 0$.

The nature of the critical points can now be determined by linearizing $q'$ and $z'$
around the critical points. The eigenvalues of the system are given by
\begin{eqnarray}
\label{eigenv}
m^2 &=&  q_c \frac{d \Delta}{dq}(q_c)
\left[1 \mp y_c^2 \pm (w_{\rm cl}-1) w_c^2 \right] \pm y_c^2(6-\Delta_c) \nonumber \\
&~&\mp \frac{1}{2} w_c^2 \left[w_{\rm cl}(\Delta_c-3)-1 \right](\Delta_c-3)(w_{\rm cl}-1)
\,.
\end{eqnarray}

Stable solutions exist when $m^2 < 0$. Substituting for
$y_c$ in Eq.~(\ref{eigenv}) using Eq.~(\ref{critical}) we observe that the eigenvalues are regular everywhere except at $q_c = 0.835$ (where $\Delta = 6$) and $q_c = 1$. These are the points where
$y_c$ and $d\Delta/dq$ blow up and change sign, respectively. Therefore,
a transition from a stable critical point ($m^2 < 0$) to a saddle point ($m^2 > 0$)
is expected at these values of $q_c$.
Of course, transitions may occur
at additional points. We depict these transitions in Fig.~\ref{sol1}
by representing the stable points with solid lines and the saddle points with dashed lines. The classical equation of state is assumed $w_{\rm cl} = 1/3$. We will always use this equation of state when dealing with specific examples in the figures, but keep the analysis general in the analytical derivations.
%

\section{Phase space trajectories}
In this section we will turn our attention to evaluate the trajectories of the phase space. The first point to keep in mind is that the kinetic energy of the field must
always be definite positive, i.e. $x^2 \ge 0$. Using the Friedmann constrain we can rewrite this condition as $ z^2 \ge \pm y^2 \pm w^2 -1$. On the other hand, since $z^2 \ge 0$ we have that in the regions of the phase space where $\pm y^2 \pm w^2 -1 \le 0$ the latter condition on $z^2$ is automatically satisfied. Conversely, in the regions where $\pm y^2 \pm w^2 -1 > 0$, there are regions of {\it exclusion} for $Ha_* = z/\sqrt{q}$, with upper and lower bounds defined as
\begin{equation}
\label{regions}
-b < H a_* < b \,,
\end{equation}
where $b = (\pm y^2 \pm w^2 -1)^{1/2}/\sqrt{q}$.
By inspecting these constraints we can have an idea of how many regions of exclusion to expect given a pair $(v,r)$. First we recall from Eq.~(\ref{y2w2}) that $y^2$ increases linearly with $q$.
We note that $\rho_{\rm b} \propto q^{6 w_{\rm cl} - 3/2}$ diverges as
$q \rightarrow 0$ when $w_{\rm cl} < 1/4$, hence, we will not consider these
cases from now on. Since, $w^2 \propto q^{6w_{\rm cl}-1/2}$ in the semi classical phase
it follows that
$w^2$ increases with $q$ in the semi classical phase.
Moreover, $w^2$ decays away as $w^2 \propto q^{-(1+3w_{\rm cl})/2}$ for $q > 1$.
In general there are four cases to consider. They correspond to the various combinations of signs before $y^2$ and $w^2$ in the definition of $b$. We represent those combinations by the pair $(i,j)$ where the first element represents the sign before $y^2$ and the second element, the sign before $w^2$.

(i) When we have the pair $(+,+)$, we can have up to two regions of exclusion which merge as either $v$ or $r$ increase. One region appears at small but non-vanishing $q$ and the second extends from small $q$ to infinity.

(ii) When we have $(+,-)$, again we can have up to two regions. This time, they merge as the ratio $|v/r|$ increases. The regions are qualitatively like the ones in (i).

(iii) When the choice of signs is $(-,+)$, there is only one possible region of exclusion which exists for small non-vanishing $q$. The area of the region increases with increasing ratio $|r/v|$.

(iv) Finally for the combination $(-,-)$, there is no region of exclusion and the whole phase space is available.

\begin{figure}[!t]
\includegraphics[width = 8.5cm]{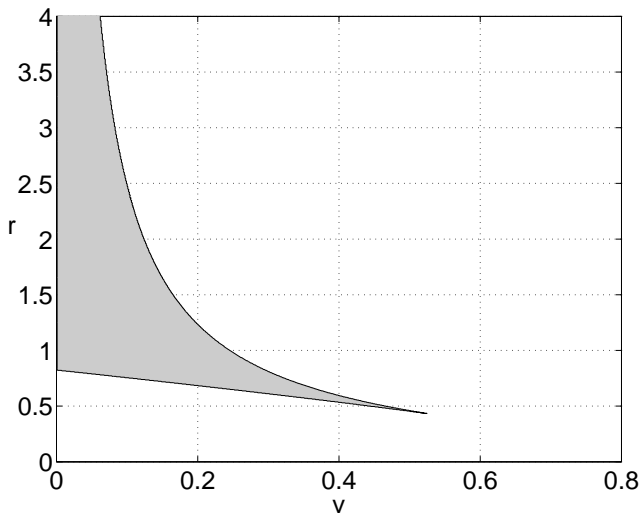}
\caption{\label{excl1} Areas of the parameter space $(v,r)$ where there exists one region of exclusion (unshaded area) and where there are two exclusion regions (shaded area). Regions of exclusion are regions in the phase space where the kinetic energy of the field is negative.}
\end{figure}

This is a qualitatively description that can be complemented by a quantitative analysis shown in Fig.~\ref{excl1}.
The shaded area corresponds to choices of $(v,r)$ that lead to two regions of exclusion in the phase space of the system.

\begin{figure}[!t]
\includegraphics[width = 4.2cm]{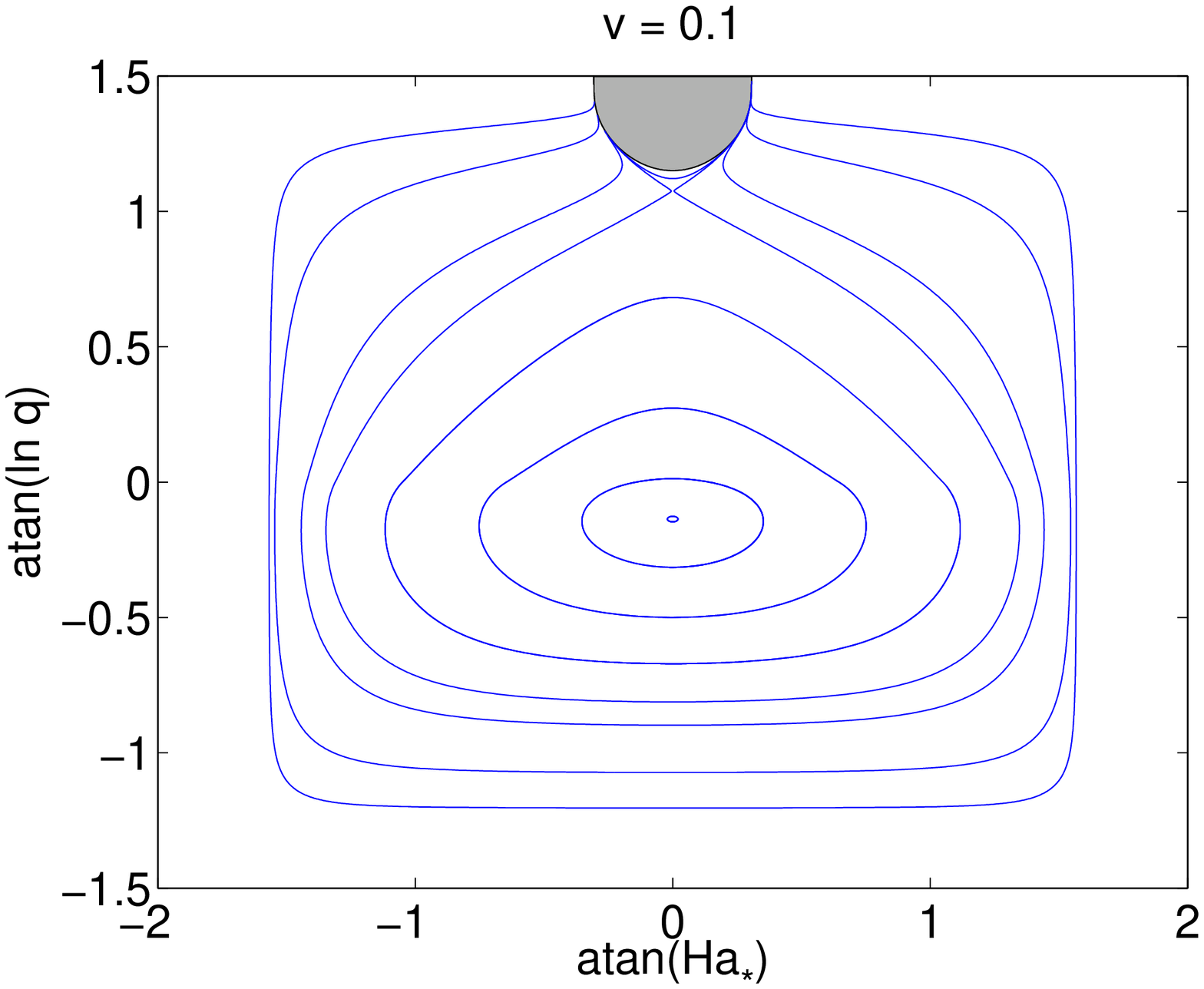}
\includegraphics[width = 4.2cm]{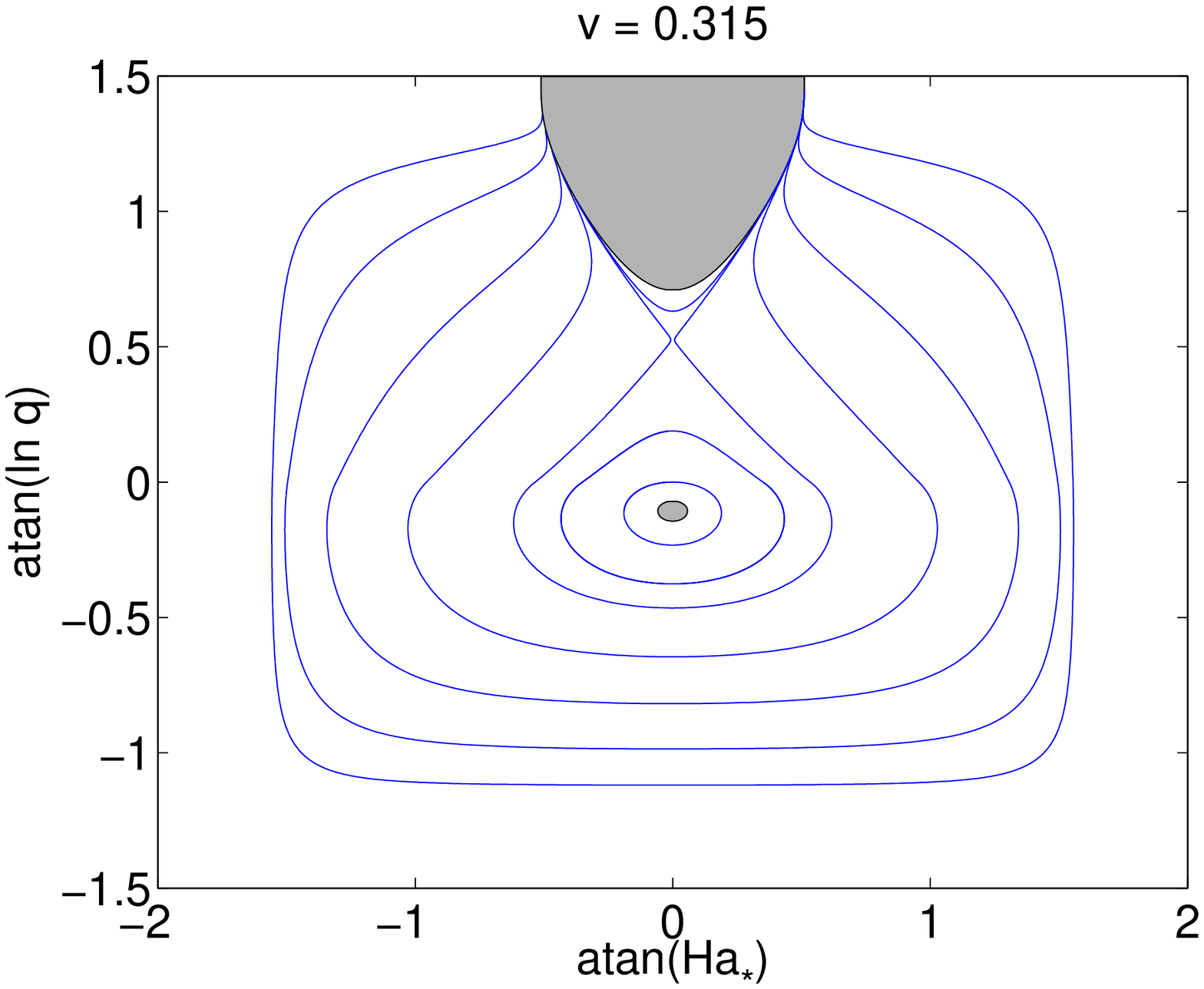}
\includegraphics[width = 4.2cm]{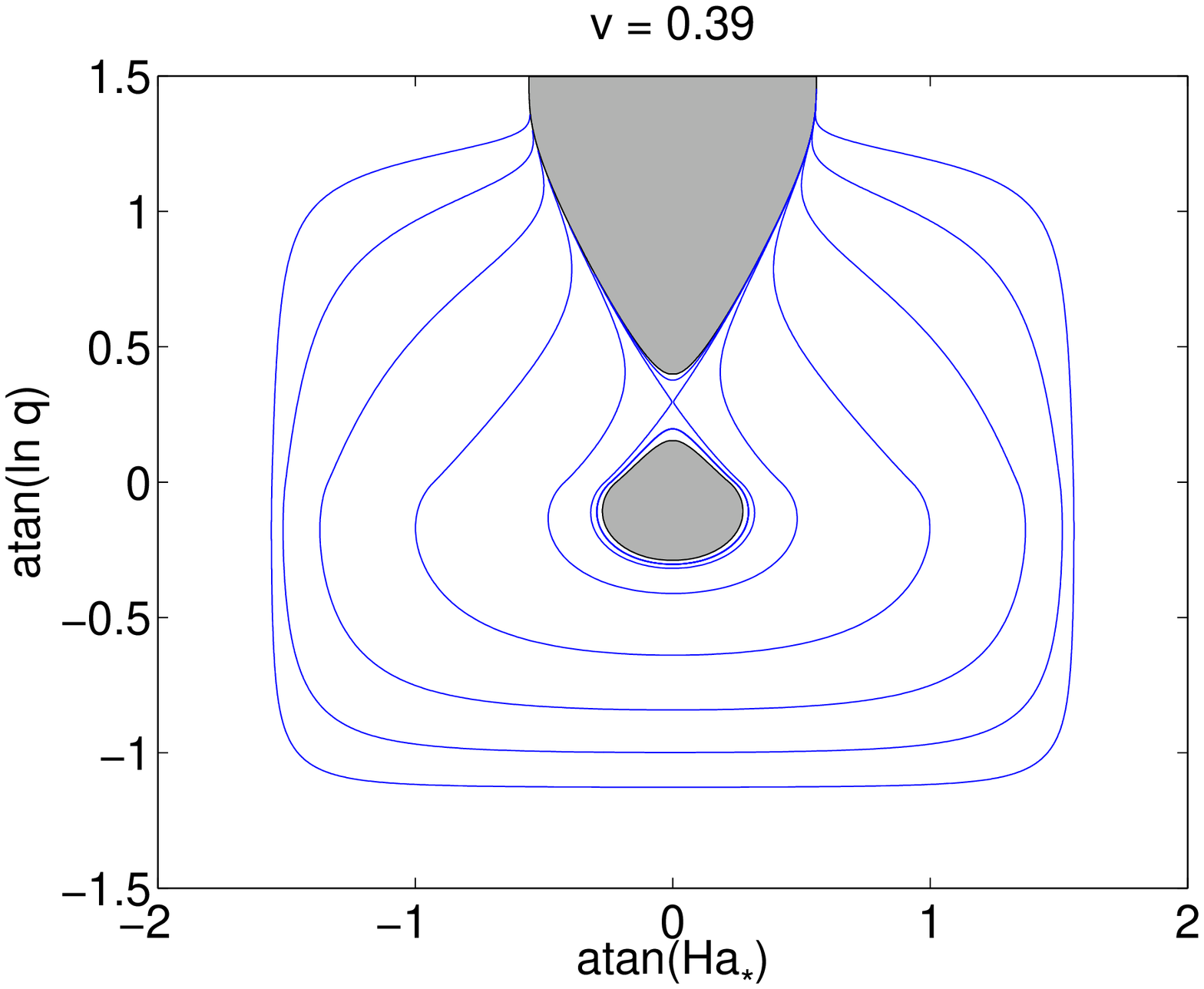}
\includegraphics[width = 4.2cm]{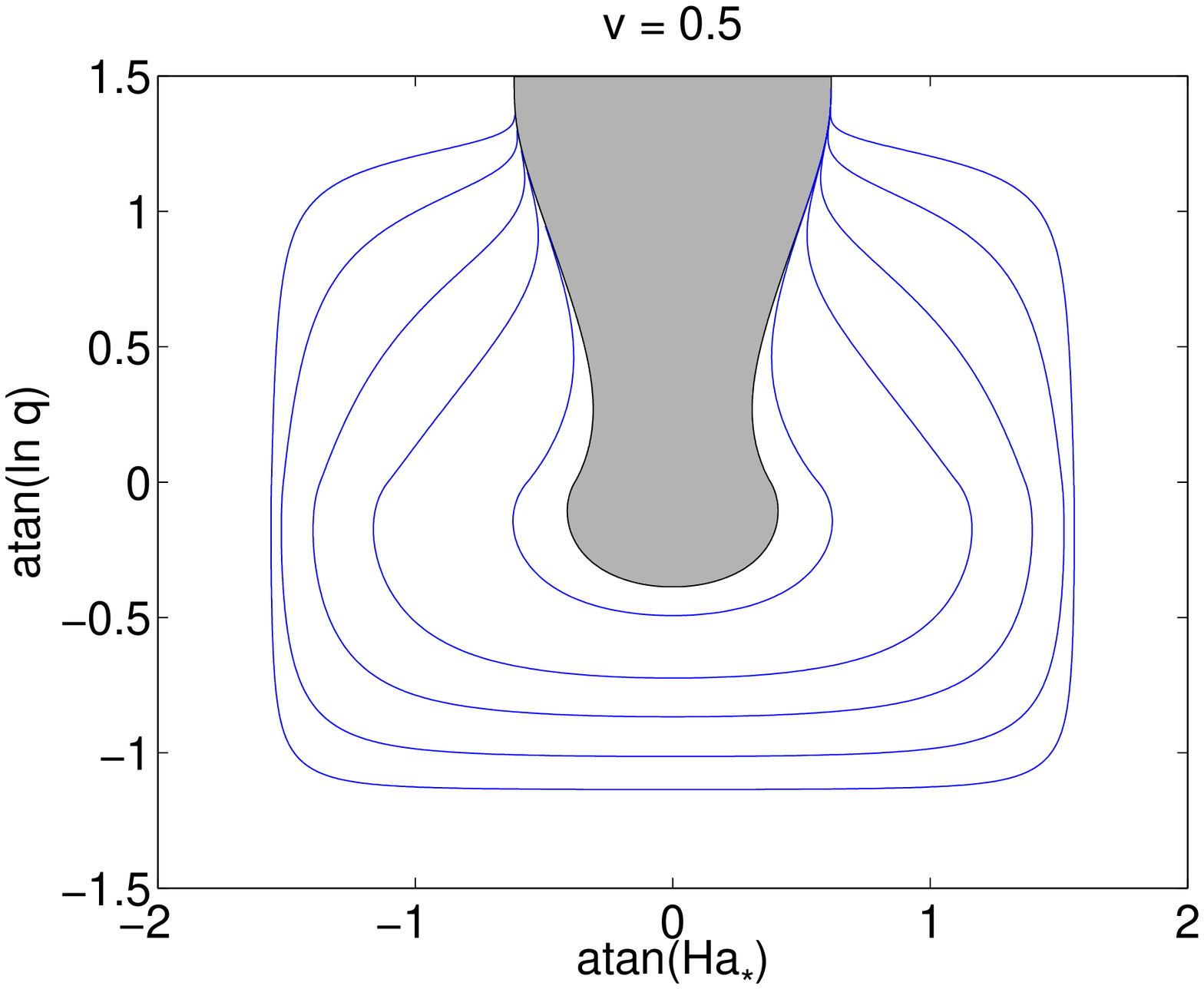}
\caption{\label{orbitv} Illustrating the phase space trajectories (solid lines)
and regions of exclusion (shaded areas) for $r = 0.6$ and varying $v$. The trajectories evolve counter clockwise.}
\end{figure}

In Fig.\ref{orbitv} we illustrate how the areas of exclusion and trajectories evolve if we
fix the value of $r$ and increase the potential. Indeed, Fig.~\ref{orbitv} corroborates the information given in Fig.~\ref{excl1}. More specifically, setting $r = 0.6$,
we verify that up to $v \approx 0.31$ there is only one region of exclusion, however, a second region appears above this value. This region increases in area as $v$ is further increased and merges with the first region for $v \approx 0.4$.

\begin{figure}[!t]
\includegraphics[width = 8.5cm]{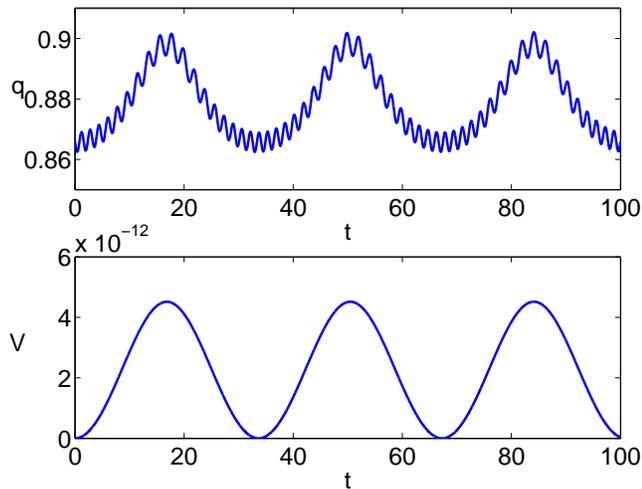}
\caption{\label{expl1} Evolution of $q$ and the potential $V$, with time. The initial conditions are $\phi_{\rm init} = 0$, $H_{\rm init} = 0$, $q_{init} = 0.866$,
$j = 1.94 \times 10^{11}$ and $r = 0.6$. The $V$ axis is labeled in Planck units.}
\end{figure}

Let us now look at a specific example in the context of a realistic varying potential.
We assume that the field is initially $\phi_{\rm init} = 0$ and does not go any further
than $\phi = 3\mpl$. Again we choose $r = 0.6$. From Fig.~\ref{sol1} we have that for such a value of $r$, the maximum value of $v$ allowed by the kinetic constraint is
$v_{\rm max} = 0.31$. For a quadratic potential
$V = m^2 \phi^2/2$ with $m = 10^{-6} \mpl$, we have in order to concretize the point of maximum displacement, $\phi = 3\mpl$, that the quantization parameter must be
$j = 1.94 \times 10^{11}$. Moreover, we can also read from Fig.~\ref{sol1}
that when $V = 0$, $q_c \approx 0.866$.
Choosing this value for $q_{\rm init}$ along with the other parameters,
we obtain the evolution shown in Fig.~\ref{expl1}. Indeed we see that the potential and the scale factor are oscillating in two fashions. One around the instantaneous critical point and the second between the critical points corresponding to $V = 0$ and
$V = 3 v_{\rm max}/\kappa^2 a_*^2$. We can understand this behavior by returning to Fig.~\ref{excl1}.
The evolution starts at $v = 0$ and develops along the line $r = 0.6$.
As the potential increases, a second region of exclusion is generated at the position of the critical point when $v = 0.31$. This region increases as the value of the potential increases. In fact, it can become as large as the orbit of the trajectory, in which case, by the definition of the regions of exclusion, $x^2 = 0$. This means that the field has
reached the point of maximum displacement and turns around.
The potential decreases and so does the area of the bounded region of exclusion. Once the field passes through zero and changes sign, the evolution mimics the evolution on the opposite side of the potential i.e. the field enters a cyclic regime.

\section{Consequences for inflation and ``graceful entrance''}
\label{graceful}
We have seen in the previous section that, in some cases, as the potential increases, an exclusion region appears in the phase space which can cause the field to loose all its kinetic energy and therefore to turn around in the potential, however, in a regime of static solutions $\dot{a} = \ddot{a} = 0$, not in an accelerated expansion,
$\ddot{a} > 0$, as required for inflation.
Looking at Fig.~\ref{excl1}, we see that provided $0< r < 0.38$,
no second region is formed.
The upper bound corresponds to the value of $r$ above which $x^2(q=1)$ in Eq.~(\ref{kinetic})
changes sign, or in other words, is the lowest value of $r$ 
for which the corresponding solid line and dashed line (like the ones in Fig.~\ref{sol1}) do not touch at $q=1$.
Hence, the area of parameter space $0< r < 0.38$, allows in principle, a smooth transition between a regime where the field is being pushed up the potential to a phase where this solution becomes unstable and the field slow rolls back down the potential thus making the Universe to inflate in the very same spirit of
Refs.~\cite{Lidsey:2004ef,Mulryne:2004va,Mulryne:2005ef}.

In this article we are primarily interested in those situations in which the background fluid is initially dominating the evolution of the Universe, and the curvature term is  important only at the values of the scale factor where the bounces occur. More specifically, neglecting the contribution of the scalar field, we require at the bounces $\kappa^2 \rho_{\rm b}(a = a_{1,2})/3 \approx a_{1,2}^{-2}$ (or equivalently $\pm w_{1,2}^2 = 1$), which leads to the constraint
\begin{equation}
\label{weq1}
r \, D_{1,2}^{w_{\rm cl}} \, q_{1,2}^{-(1+3w_{\rm cl})/2}  = 1 \,.
\end{equation}
This equation has two real roots ($q_1$ and $q_2$)
provided $r > 0.82$. It is important to
stress that this value lies outside the bound $0 < r < 0.38$ that avoids the generation of a second exclusion region. This means that {\it it is not trivial to implement a mechanism akin to the one discussed in Refs.~\cite{Lidsey:2004ef,Mulryne:2004va,Mulryne:2005ef}
when a
dominant background fluid is present}. In fact, we can only expect a transition from an oscillating Universe into an inflationary one by admitting that the amplitudes of the oscillations of the scale factor are sufficiently large
such that the trajectories in the phase space touch the separatrix (where the trajectories appear to intersect) before the diameter of the second exclusion region equals the diameter of the trajectory.

What is therefore the correct values of initial parameters that deliver the desired evolution? Admitting that initially the evolution starts with $V = 0$ and $H = 0$, we have $x_{\rm init}^2 + w_{\rm init}^2 = 1$. On the other hand,
when the field turns around in the potential to slow roll back down in the last cycle,
very near when the
two exclusion regions merge, one has
$y_{\rm merge}^2 + w_{\rm merge}^2 = 1$. Further admitting that $q_1 = q_{\rm init} \approx q_{\rm final}$ and defining the useful quantity
\begin{equation}
\label{defalpha}
\alpha \equiv \frac{\dot{\phi}_{\rm init}^2}{D_{\rm init} \, {\rho_{\rm b}}_{\rm init}} \,,
\end{equation}
that relates the kinetic energy of the field to the energy density of the background,
using Eq.~(\ref{weq1}) and requiring $x_{\rm init}^2 > y_{\rm merge}^2$ we find that
\begin{equation}
\alpha > 2 \, q_{\rm init} v_{\rm merge} \,.
\end{equation}
It is concluded that the ratio of kinetic energy to the background's energy density is bound from below by a quantity that is independent of $j$. More specifically, we realize that it increases for increasing $q_{\rm init}$. Indeed, from Eq.~(\ref{weq1}) when $q_1$ ($= q_{\rm init}$) increases, $r$ decreases, and
Fig.~\ref{excl1}, implies that $v_{\rm merge}$ increases.
For example, if $q_{\rm init} = 4 \times 10^{-4}$, the two region of exclusion merge for $v \approx 6 \times 10^{-7}$ which results into the limit $\alpha > 10^{-10}$. It is however worth pointing out that if $q_{\rm init}$ is small the consistency condition $H a_i < 1$ is violated as can be seen from Fig.~\ref{jmin}. This constraint simply states that the Hubble radius must be larger than the limiting scale of the theory.
On the other hand, for large $q_{\rm init}$, the ratio $\alpha$ becomes close to unity meaning that the contribution of the field must be important.

\begin{figure}[!t]
\includegraphics[width = 8.5cm]{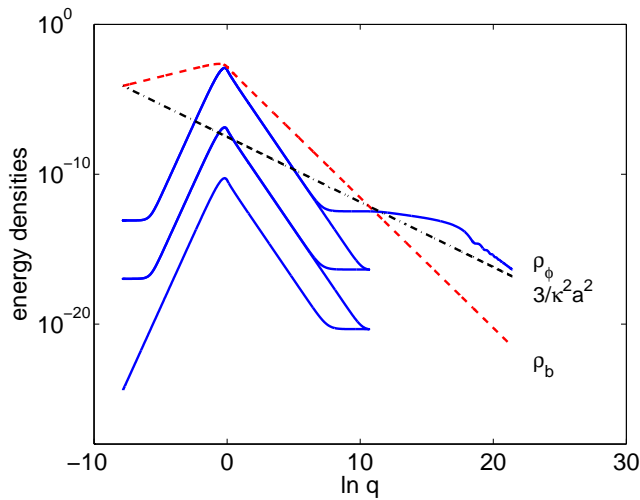}
\caption{\label{rhos} Evolution of the energy densities with the scale factor. The scalar field is represented by the solid line, the background fluid by the dashed line and the curvature term by the dash-dotted line. We have used $\alpha = 10^{-20}$,
$\phi_{\rm init} = 0$,
$q_{\rm init} = 4 \times 10^{-4}$ and $j = 9 \times 10^7$ for a quadratic potential $V = m^2 \phi^2/2$ with $m = 10^{-6} \mpl$. The vertical axis is labeled in Planck units.}
\end{figure}

  In the remainder of this paper we deal with an alternative mechanism where the energy density of the scalar field, though initially subdominant, builds its way up during the numerous cycles of oscillation of the Universe until it becomes dominant delivering an accelerated expansion. We show in Fig.~\ref{rhos} an example of such an evolution. The example yields only a few $e$-folds of inflation, for illustrative purposes, but as we shall see below, there are regions of parameter space that provide a successful inflationary scenario with more than $60$ $e$-folds of expansion.
The evolution of the field starts in the semi-classical phase,
$\ln q_{\rm init} = -7.8$, at the bottom of the potential with a kinetic energy that is twenty orders of magnitude smaller than the energy density of the background fluid. The anti-frictional term in the equation of motion of the scalar field accelerates the field up the potential as the Universe expands. The energy density of the field evolves with an effective equation of state $w_\phi = -4$. Once in the classical phase ($a> a_*$) this term turns into a frictional component forcing the field to slow down. The equation of state, therefore, approaches $w_\phi = 1$. In the case of Fig.~\ref{rhos}, the field's kinetic energy decreases below the potential energy at $\ln q_f \approx 7.8$ and comes to a stand and turns around in the potential ($w_\phi = -1$)
when $\ln q_t \approx 9.7$ (indexes ``$f$'' and ``$t$'' stand for ``freeze'' and ``turn around'', respectively). On the way down the potential, the field gains again
kinetic energy which is enhanced once the Universe recollapses (as in the classical phase the frictional term becomes anti-frictional when the Hubble ratio is negative).
The field is again slowed down as soon as the scale factor enters the semi-classical phase during the collapse. When the kinetic energy becomes smaller than the potential, the
equation of state again approaches $-1$. In this example, the kinetic energy decreases but it does not vanishes in the semi-classical phase; hence, when the Universe goes through the bounce starting a new cycle, the field keeps moving in the same direction. The subsequent evolution follows closely the previous description for a number of oscillations of the Universe.
The remarkable feature about the evolution is that the field gains energy on each cycle thus allowing it to be displaced by a larger amount from one cycle to the next. When the contribution of the energy density of the field becomes important,
the field can slow roll and the Universe naturally starts
an inflationary evolution say, in a "graceful entrance".

\section{Parameter space}
  At this point we need to establish which set of the parameter space leads to a viable model of inflation rather than to an infinite series of oscillations of the Universe. Let us start by calculating by how much the scalar field is displaced in half a cycle of the Universe. We assume that the initial value of the scale factor is deep into the semi-classical phase and that the potential is initially negligible.
  From the equation of motion of the scalar field, neglecting the contribution of the potential, we obtain
\begin{equation}
H \frac{d \phi}{d \ln a} = \dot{\phi}_{\rm init} \left(\frac{a}{a_{\rm init}} \right)^{-3} \frac{D}{D_{\rm init}} \,.
\end{equation}
Using $H^2 \approx \kappa^2 \rho_{\rm b}/3$ (this is obviously not true at the bounces but it is a good approximation within the extremes of the scale factor) and integrating one gets
\begin{eqnarray}
\label{phif}
\phi_f &=& \phi_{\rm init} + \frac{2}{3}
\frac{q_{\rm init}^{3/2} \, a_*}{ D_{\rm init} \sqrt{r}} \,
\frac{\dot{\phi}_{\rm init}}{(9-4 w_{\rm cl})(1-w_{\rm cl})}
\nonumber \\
&~& \times \left\{ (10-5w_{\rm cl}) \left(\frac{12}{7}\right)^{3(1-w_{\rm cl})/5}
    \right. \nonumber \\
&~&  - (1-w_{\rm cl}) \left(\frac{12}{7}\right)^{3(1-w_{\rm cl})}
    q_{\rm init}^{(27- 12 w_{\rm cl})/4} \nonumber \\
&~& \left. -(9-4w_{\rm cl}) \, q_f^{-3(1-w_{\rm cl})/4} \right\} \,,
\end{eqnarray}
where $\phi_f$ and $q_f$ correspond to the value of the field and $q$ when the field freezes in the potential, respectively.
To perform this integration we have considered two epochs of evolution. In the semi-classical phase, between $a_{\rm init}$ and $a_S$ the function $D$ takes the asymptotic form $D_{\rm approx} = (12/7)^6 \, q^{15/2}$. In the classical phase, when $a > a_S$ we take $D = 1$. The quantity $a_S$ is defined as the value of the scale factor where
$D_{\rm approx} = 1$. We can estimate the value of $q_f$ by using the expression
\begin{equation}
\label{phidott}
\dot{\phi}_f = \dot{\phi}_{\rm init} \left(\frac{7}{12}\right)^6 \,
\frac{1}{q_{\rm init}^6 \, q_f^{3/2}} \,,
\end{equation}
which follows from using the first integral of the equation of motion of the field along with the asymptotic form $D_{\rm approx}$.
Furthermore, a freezing point can only be identified provided
the potential becomes at least as large as the kinetic energy, such that,
$m^2 \phi_f^2/2 \approx \dot{\phi}_f^2$, for a quadratic potential.
Using Eq.~(\ref{phidott}) we obtain
\begin{equation}
\label{qt}
q_f^3 = 2 \left(\frac{7}{12}\right)^{12} \frac{\dot{\phi}_{\rm init}^2}{m^2 \phi_f^2} \, \frac{1}{q_{\rm init}^{12}} \,.
\end{equation}
Using Eq.~(\ref{defalpha}) we can write the initial kinetic energy as
\begin{equation}
\label{dotphii}
\dot{\phi}_{\rm init}^2 = 3 \, \alpha \,r \, \frac{1}{\kappa^2 a_*^2} \, q_{\rm init}^{-3(1+w_{\rm cl})/2} \, D_{\rm init}^{1+w_{\rm cl}} \,.
\end{equation}
Inserting this expression into Eq.~(\ref{phif}),
the multiplicative factor before the curly brackets reads,
\begin{equation}
\frac{2}{\kappa} \, \sqrt{\frac{\alpha}{3}} \, D_{\rm init}^{(w_{\rm cl}-1)/2} \,
q_{\rm init}^{3(1-w_{\rm cl})/4} \,,
\end{equation}
which is explicitly independent of $a_*$.
Moreover, when $\dot{\phi}_{\rm init}^2 \gg \dot{\phi}_f^2 \approx m^2 \phi_f^2/2$ the $q_f$ dependent term  in Eq.~(\ref{phif}) can be neglected and then the displacement $\phi_f$ is completely
independent of the value of $j$ for a given $q_{\rm init}$ and $\alpha$.
From Eq.~(\ref{dotphii}) and substituting for $r$ using Eq.~(\ref{defrv}) we get $\dot{\phi}_{\rm init}^2 \propto a_*^{-3(1+w_{\rm cl})}$, hence we expect $\phi_f$ to become independent of $j$ for small values of this parameter.

\begin{figure}[!t]
\includegraphics[width = 8.5cm]{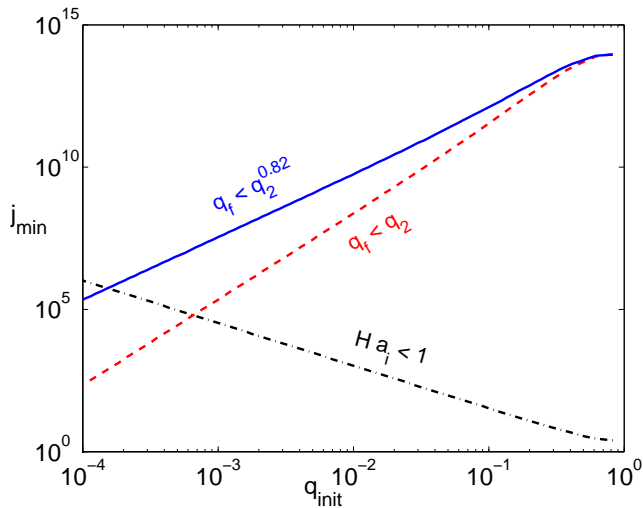}
\caption{\label{jmin} The dashed line represents the lower limit on $j$ resulting from requiring that the field freezes before the bounce at $q_2$. This is estimated using Eqs.~(\ref{phif}) and (\ref{phidott}).
The solid line represents the borderline between an evolution that pushes the field up the potential in successive steps (below the curve) and an evolution where the field is displaced from the initial position stops before the bounce, turns around and is displaced again in the opposite direction in the collapsing phase (above the curve). This curve was obtained numerically and fits to $q_f = q_2^{0.82}$ accurately. The dash-dotted line gives a lower bound on the values of $j$ that satisfy the consistency condition $H a_i < 1$.}
\end{figure}

A necessary condition for the field to turn around is that $q_f < q_2$, i.e., it must freeze before the Universe recollapses.
This constraint results into a lower bound on the value of the quantization parameter $j$ with respect to the initial value of $q_{\rm init}$. We draw this bound in Fig.~\ref{jmin}, dashed line.
This condition might not be sufficient to guaranty that the field stops and reverses its direction, as the kinetic energy can decrease below the value of the potential but without vanishing. In this latter case the field keeps moving up the potential in a series of steps before the turn around occurs (such cases were discussed in the first part of Section \ref{graceful}, an example is illustrated in Fig.~\ref{expl1}, and do not result into an inflationary expansion, as we have seen). Understandably, $\ln q_f$
must be a fraction of the value of $\ln q_2$. By numerically integrating the equations of motion it can be verified that this fraction is nearly a constant and its value is
$0.82$. In Fig.\ref{jmin} we show by a solid line the lower limit on $j$ for which the field does changes direction on each cycle corresponding to $q_f < q_2^{0.82}$.

We also draw in Fig.~\ref{jmin} the limit on $j$ imposed by the consistency condition $H a_i < 1$. Using the approximation that the maximum of the background energy density occurs when $q = 1$ and $D = 1$, this condition is equivalent to $j > 3\,r$. Using
$q_1 = q_{\rm init}$ in Eq.~(\ref{weq1}) and the asymptotic form $D_{\rm approx}$, we get that deep in the semi-classical phase the limit on $j$ can be explicitly written in terms of the initial value $q_{\rm init}$ as
\begin{equation}
j > 3 \left(\frac{12}{7}\right)^{-6w_{\rm cl}}\, q_{\rm init}^{(1-12 w_{\rm cl})/2} \,.
\end{equation}
From Fig.~\ref{jmin} it is concluded that below $q_{\rm init} = 2 \times 10^{-4}$ the consistency constraint establishes the most important lower limit on $j$.

\begin{figure}[!t]
\includegraphics[width = 8.5cm]{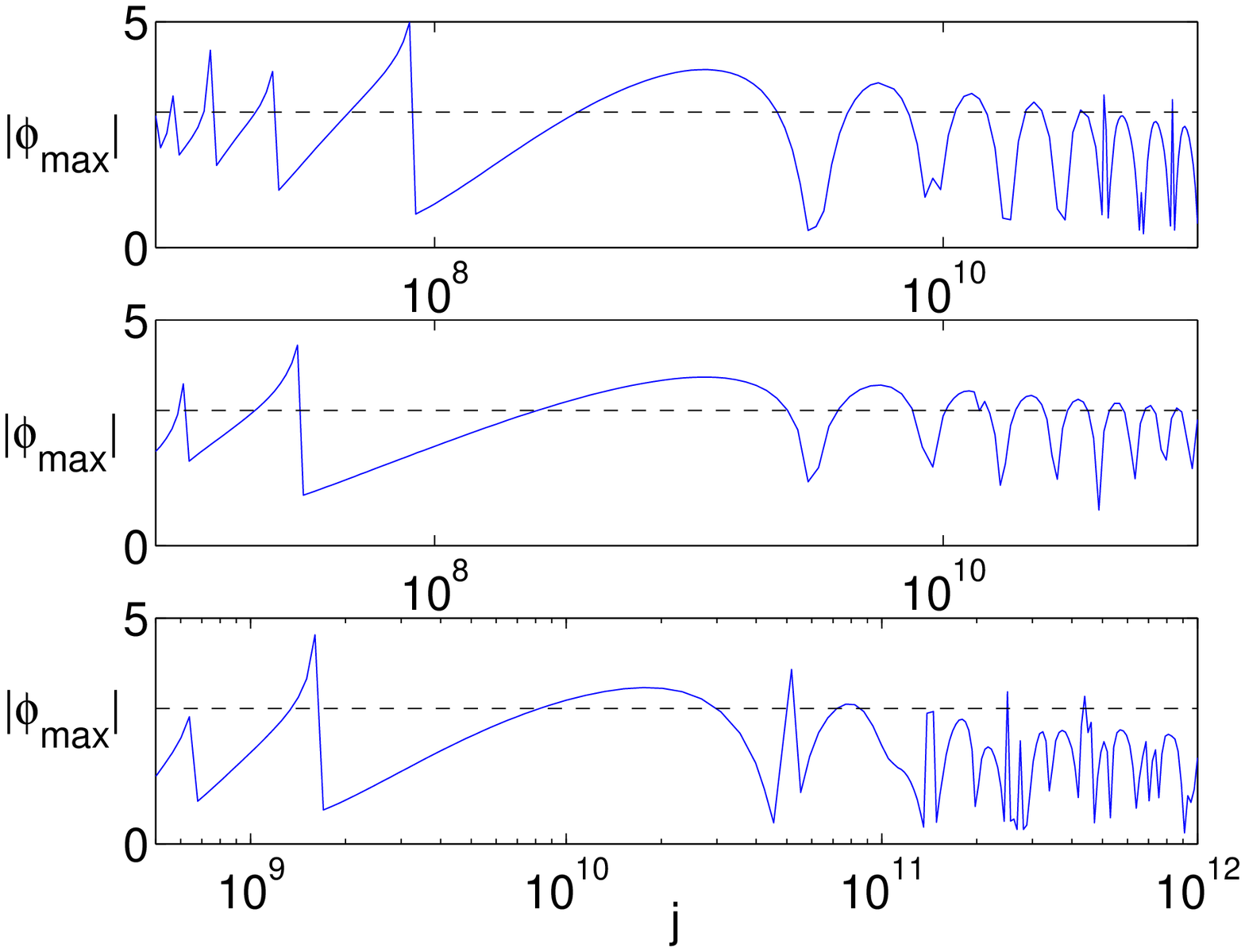}
\caption{\label{phimax} Maximum achieved value of the field (in Planck units)
with respect to the quantization parameter $j$ for three different sets of initial conditions.
Upper panel:  $q_{\rm init} = 4 \times 10^{-4}$ and $\alpha = 10^{-20}$;
middle panel: $q_{\rm init} = 4 \times 10^{-4}$ and $\alpha = 2 \times 10^{-15}$;
lower panel: $q_{\rm init} = 2.5 \times 10^{-3}$ and $\alpha = 2 \times 10^{-15}$.
Dashed lines represent $\phi = 3 \mpl$.
The vertical axes are labeled in Planck units.}
\end{figure}

\begin{figure}[!t]
\includegraphics[width = 8.5cm]{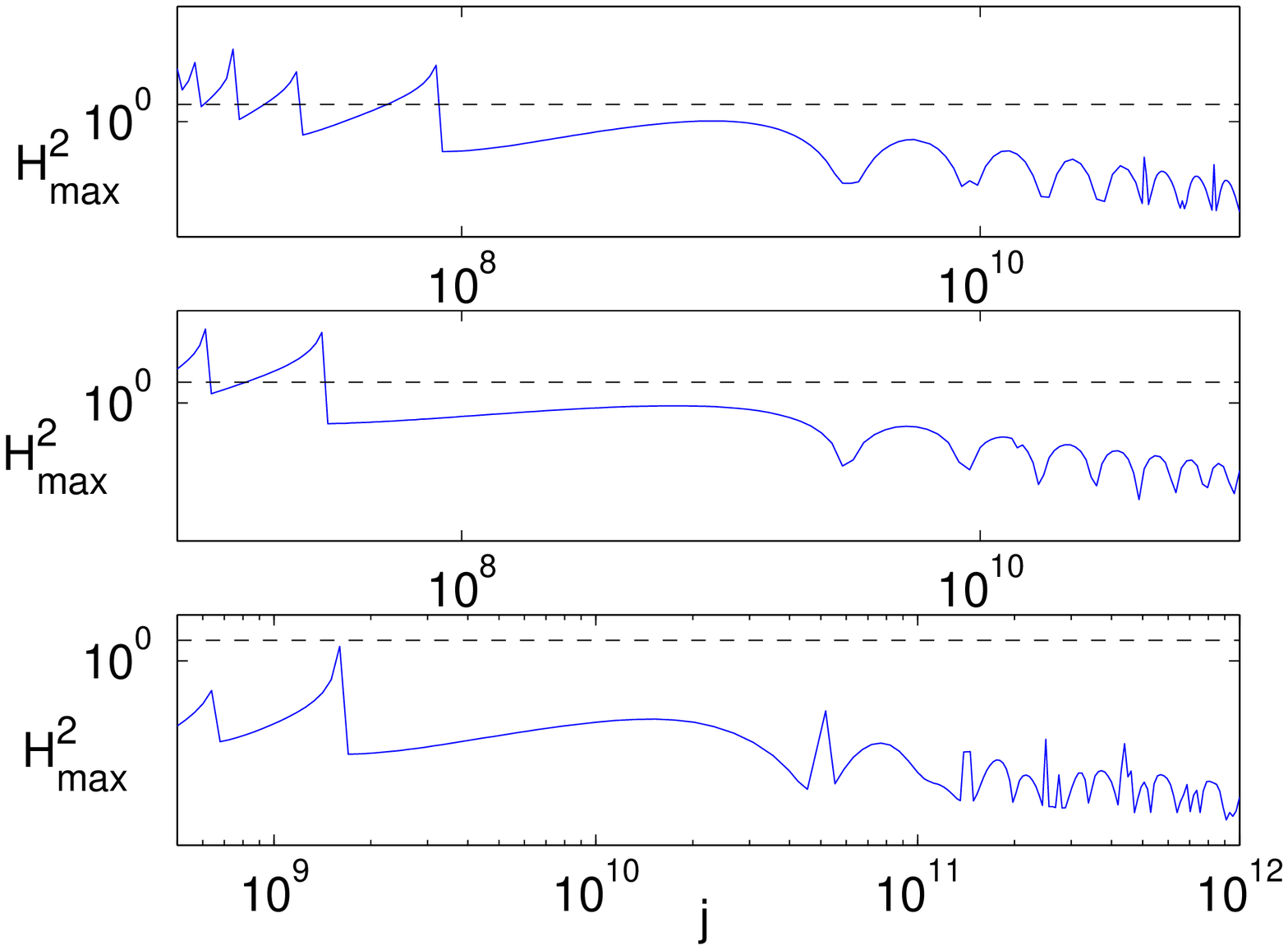}
\caption{\label{H2max} Maximum achieved value of the Hubble ratio (in Planck units)
 with respect to the quantization parameter $j$ for three different sets of initial conditions.
Upper panel:  $q_{\rm init} = 4 \times 10^{-4}$ and $\alpha = 10^{-20}$;
middle panel: $q_{\rm init} = 4 \times 10^{-4}$ and $\alpha = 2 \times 10^{-15}$;
lower panel: $q_{\rm init} = 2.5 \times 10^{-3}$ and $\alpha = 2 \times 10^{-15}$.
Dashed lines represent the consistency constraint $H a_i < 1$.
The vertical axes are labeled in Planck units.}
\end{figure}

Finally we can evaluate how efficient the mechanism is.
By numerically integrating the equations of motion it is possible to determine how high can the field move in the potential before it rules the dynamics of the Universe.
More specifically, we have evolved the system up to $q_{\rm end} = 1+ (\ln q_2) /2$, i.e. until the scale factor
is one e-fold larger than its maximum value at the bounces, when the background fluid is dominant. We have assumed a quadratic potential $V = m^2 \phi^2/2$ with
$m = 10^{-6} \mpl$.
Figures \ref{phimax} and \ref{H2max} show the maximum value of the field and the maximum value of the Hubble ratio, respectively, with respect to the quantization parameter $j$. In all cases, the evolution starts from $\phi_{\rm init} = 0$.


Important information can immediately be extracted from these figures. First, we note from Fig.~\ref{H2max} that the consistency condition $ H a_i < 1$ (which is equivalent to $H^2 < 7.7$ in Planck units) is only satisfied provided $j > 8 \times 10^{7}$, $j > 3 \times 10^7$ for the first and second examples illustrated.
Above these limits there are values of $j$ for which the maximum value of the field is above three Planck units, hence providing a viable inflationary scenario with more than 60 $e$-folds of accelerated expansion. We note, however, that by increasing $q_{\rm init}$ the maximum achieved value of $\phi$ decreases, as it is seen by comparing the middle and bottom panels of Fig.~(\ref{phimax}).

A remarkable feature about these figures is that they present a
tooth wood saw bladelike structure for low $j$ values.
This can easily be understood qualitatively. Indeed, from Eqs.~(\ref{qt}), (\ref{dotphii}) and substituting for $r$ using Eq.~(\ref{defrv}), we see that the value of $q_f$
decreases for increasing $j$ as $q_f \propto a_*^{-1-w_{\rm cl}}$ (under the approximation that $\phi_f$ does not depend on $j$, which is accurate for $\dot{\phi}_{\rm init}^2 \gg \dot{\phi}_f^2$ as we have seen before). This means that in a Universe where $j$ is large, the field freezes and consequently starts rolling down when the scale factor is smaller than in a Universe with low $j$.
The field is accelerated during the expansion phase
between $q_t$ and $q_2$ (even though the $\dot{\phi}$ term in the equation of motion is a frictional one), during the collapse between $q_2$ and $q \approx 1$, and again during the expansionary phase when $q < 1$. But since the field starts moving earlier in a Universe for larger $j$, the net effect is that it is pushed further up the potential in this case.
In other words, the field gains kinetic energy faster if $j$ is large,
which results into a larger displacement of the field from its initial position.
This explains why the maximum value of the field increases with increasing $j$
in Fig.~\ref{phimax}. The reason why the maximum value of the field suddenly drops by increasing $j$ is understood by realizing that for a critical value $j_c$ there is one less cycle than if $j \lesssim j_c$, thus preventing the field to move as high in the potential. As we keep increasing $j$, the description made above follows, now with one less cycle of
expansion and recollapse.

Along the same lines, a similar oscillatory structure is expected by varying $\alpha$. We have seen in Fig~\ref{rhos} that for a given $j$ the energy density of the field increases on each cycle by a nearly constant amount. It can be verified that if
$\alpha$ was instead $10^{-16}$ ($10^{-24}$)
the evolution would nearly reproduce the one in
Fig.~\ref{rhos} with the only exception that there would be one less (more) cycle.

Another prominent feature in the figures are the sharp peaks giving place to smooth extremes above some value of $j$. Indeed, when $j$ becomes sufficiently large, the field has enough time to increase its speed before the recollapse takes place. We have in those cases $\dot{\phi}^2 > V$ at $q = q_2$. Therefore, the field's equation of state is considerably larger than $w_\phi = -1$ at this point. This behavior translates into a similar effect of the one of having a slightly smaller $j$, i.e. the ratio of energy densities from one cycle to the next is smaller than expected had the potential energy remained always dominant. As $j$ becomes even larger, the field is able to perform a few oscillations around the value at the minimum of the potential before the collapse occurs.
  To summarize, the effect of increasing $j$ is compensated by the fact that the field loses most of its potential energy, thus the number of cycles of the Universe before the field becomes dominant does not decrease anymore as $j$ is increased.

\section{Discussion}
In this article we have looked upon the dynamics of a scalar field evolving
in a closed Universe where a background fluid is present in the context of loop quantum cosmology.
In the first part we have searched for static solutions of the system and
evaluated their stability admitting a constant potential. We have discovered that, for a constant value of $\rho_0$ and $j$, as the potential is increased, a second region of exclusion (where the kinetic energy is negative) is generated in the phase space.
We argued that the occurrence of such a region makes it difficult to implement a similar mechanism to the one studied in Refs.~\cite{Lidsey:2004ef,Mulryne:2004va,Mulryne:2005ef} where the field moves up the potential in a series of steps resulting from the cyclic expanding and contracting phases of the Universe. A viable model requires the phase space trajectories of the system to be sufficiently away from the critical point in order to avoid the exclusion region since its area grows as the potential increases. This requirement translates into a lower limit on the initial value of the kinetic energy of the field and is defined in the region below the curve labeled $q_f < q_2^{0.82}$ in Fig.~\ref{jmin}.

 In the second part of this work we have shown examples of an alternative mechanism which is defined when $q_f < q_2^{0.82}$. Qualitatively, it means that the field turns around on each cycle. The kinetic energy gained when the field is rolling down the potential in the expansionary phase, is enhanced during the collapse pushing the field fasrther up the potential on each cycle. The energy density of the field comes eventually to dominate the curvature and background contributions establishing the initial
conditions for inflation.

We have seen that
there are values of $j$ for which the maximum value of the field is above three Planck units, thus able to provide a viable inflationary evolution. It is nonetheless
observed that it is very easy to violate the Hubble bound $H a_i < 1$. This problem significantly constraints the range of parameter space for which one obtains a viable result within the limits of validity of the theory. Moreover, it
involves large values of the parameter $j$. Large values of the parameter $j$ are considered to be unnatural \cite{Bojowald:2004xq}. Also worth pointing out that the mechanism can only be implemented for potentials with a minimum.

In this work we have neglected discreteness corrections. This approach is valid as long as $\dot{a} \ll 1$ \cite{Banerjee:2005ga,Singh:2005xg}. Admitting that the maximum value of $\dot{a}$ occurs near $a_*$, we obtain that $H(a = a_*) a_* \ll 1$. Using the approximation that the maximum of the background energy density occurs when $q=1$ with $D = 1$, we get the equivalent condition $r \ll 1$. From Eq.~(\ref{weq1}) we can further extract a constraint on the minimum value of $q_{\rm init}$:
\begin{equation}
\left(\frac{12}{7}\right)^{6w_{\rm cl}} q_{\rm init}^{6w_{\rm cl}-1/2} \gg 1\,.
\end{equation}
More specifically, $q_{\rm init} > 0.5$ for $w_{\rm cl} = 1/3$, independently of the value of $j$.
This bound, only involving orders of magnitude, is more stringent than the heuristic consistency condition used so far in the literature, $H a_i < 1$. It also suggests that it is very difficult to move the field as far as $3 \mpl$ unless the initial kinetic energy of the field is at least comparable to the energy density of the background fluid. It would be interesting to investigate how the dynamics and bounds get modified once the discreteness corrections are taken into account.

Finally we would like to keep in mind that the oscillations of the field may
induce the conversion of part of its energy density into radiation thereby modifying the background's contribution in the process. The picture is therefore more complex than the one we dealt with here and deserves further study.

\begin{acknowledgments}
The author thanks David Mulryne for numerous discussions and for suggesting Fig.~\ref{orbitv}, and Parampreet Singh for comments on the manuscript.
This work is supported by the Department of Energy under
contract No. DE-FG02-94ER40823 at the University of Minnesota.
\end{acknowledgments}


\begin{thebibliography}{99}

\bibitem{Rovelli:1997yv}
  C.~Rovelli,
  Living Rev.\ Rel.\  {\bf 1}, 1 (1998)
  [arXiv:gr-qc/9710008].

\bibitem{Thiemann:2002nj}
  T.~Thiemann,
  Lect.\ Notes Phys.\  {\bf 631}, 41 (2003)
  [arXiv:gr-qc/0210094].

\bibitem{Corichi:2005bn}
  A.~Corichi,
  arXiv:gr-qc/0507038.

\bibitem{Bojowald:2002gz}
  M.~Bojowald,
  Class.\ Quant.\ Grav.\  {\bf 19}, 2717 (2002)
  [arXiv:gr-qc/0202077].

\bibitem{Bojowald:2002ny}
  M.~Bojowald,
  Class.\ Quant.\ Grav.\  {\bf 19}, 5113 (2002)
  [arXiv:gr-qc/0206053].


\bibitem{Bojowald:2002nz}
  M.~Bojowald,
  Phys.\ Rev.\ Lett.\  {\bf 89}, 261301 (2002)
  [arXiv:gr-qc/0206054].

\bibitem{Tsujikawa:2003vr}
  S.~Tsujikawa, P.~Singh and R.~Maartens,
  Class.\ Quant.\ Grav.\  {\bf 21}, 5767 (2004)
  [arXiv:astro-ph/0311015].


\bibitem{Bojowald:2004xq}
  M.~Bojowald, J.~E.~Lidsey, D.~J.~Mulryne, P.~Singh and R.~Tavakol,
  Phys.\ Rev.\ D {\bf 70}, 043530 (2004)
  [arXiv:gr-qc/0403106].

\bibitem{Lidsey:2004ef}
  J.~E.~Lidsey, D.~J.~Mulryne, N.~J.~Nunes and R.~Tavakol,
  Phys.\ Rev.\ D {\bf 70}, 063521 (2004)
  [arXiv:gr-qc/0406042].


\bibitem{Vereshchagin:2004uc}
  G.~V.~Vereshchagin,
  JCAP {\bf 0407}, 013 (2004)
  [arXiv:gr-qc/0406108].

\bibitem{Lidsey:2004uz}
  J.~E.~Lidsey,
  JCAP {\bf 0412}, 007 (2004)
  [arXiv:gr-qc/0411124].

\bibitem{Date:2004yz}
  G.~Date and G.~M.~Hossain,
  Phys.\ Rev.\ Lett.\  {\bf 94}, 011301 (2005)
  [arXiv:gr-qc/0407069].

\bibitem{Mulryne:2004va}
  D.~J.~Mulryne, N.~J.~Nunes, R.~Tavakol and J.~E.~Lidsey,
  Int.\ J.\ Mod.\ Phys.\ A {\bf 20}, 2347 (2005)
  [arXiv:gr-qc/0411125].

\bibitem{Mulryne:2005ef}
  D.~J.~Mulryne, R.~Tavakol, J.~E.~Lidsey and G.~F.~R.~Ellis,
  Phys.\ Rev.\ D {\bf 71}, 123512 (2005)
  [arXiv:astro-ph/0502589].



\bibitem{Singh:2003au}
  P.~Singh and A.~Toporensky,
  Phys.\ Rev.\ D {\bf 69}, 104008 (2004)
  [arXiv:gr-qc/0312110].

\bibitem{Date:2004fj}
  G.~Date and G.~M.~Hossain,
  Phys.\ Rev.\ Lett.\  {\bf 94}, 011302 (2005)
  [arXiv:gr-qc/0407074].

\bibitem{Bojowald:2004kt}
  M.~Bojowald, R.~Maartens and P.~Singh,
  Phys.\ Rev.\ D {\bf 70}, 083517 (2004)
  [arXiv:hep-th/0407115].

\bibitem{Singh:2005km}
  P.~Singh,
  Class.\ Quant.\ Grav.\  {\bf 22}, 4203 (2005)
  [arXiv:gr-qc/0502086].
  
  
\bibitem{Domagala:2004jt}
  M.~Domagala and J.~Lewandowski,
  Class.\ Quant.\ Grav.\  {\bf 21}, 5233 (2004)
  [arXiv:gr-qc/0407051].

\bibitem{Meissner:2004ju}
  K.~A.~Meissner,
  Class.\ Quant.\ Grav.\  {\bf 21}, 5245 (2004)
  [arXiv:gr-qc/0407052].


\bibitem{Banerjee:2005ga}
  K.~Banerjee and G.~Date,
  Class.\ Quant.\ Grav.\  {\bf 22}, 2017 (2005)
  [arXiv:gr-qc/0501102].


\bibitem{Singh:2005xg}
  P.~Singh and K.~Vandersloot,
  Phys.\ Rev.\ D {\bf 72}, 084004 (2005)
  [arXiv:gr-qc/0507029].
  
  
  
\end{thebibliography}
\end{document}